\documentstyle[sprocl]{article}

\def\Journal#1#2#3#4{{#1} {\bf #2}, #3 (#4)}

\def\be{\begin{equation}}
\def\ee{\end{equation}}
\def\bea{\begin{eqnarray}}
\def\eea{\end{eqnarray}}
\begin{document}

\title{DIFFUSE RADIO EMISSION IN THE COMA CLUSTER}

\author{L. FERETTI, G. GIOVANNINI}

\address{Istituto di Radioastronomia, Via Gobetti 101, \\40129 Bologna, Italy 
\\E-mail: lferetti@ira.bo.cnr.it, ggiovannini@astbo1.bo.cnr.it}


\maketitle\abstracts{ 
The Coma cluster is peculiar in the radio domain owing to the presence 
of cluster-wide diffuse sources: the radio halo Coma C,  
the relic source 1253+275, and the bridge between them. 
Diffuse radio sources in clusters  are a rare and poorly understood
phenomenon. 
We summarize here the properties of the diffuse radio sources in Coma
and their relation to other cluster properties. }

\section{Radio emission in Coma}

The radio emission in clusters of galaxies is due in the majority of
cases to the
individual radio emitting galaxies. The Coma cluster is indeed
characterized by the presence of 29 cluster radio galaxies, of different
optical morphological types.
Moreover, the Coma cluster is permeated by diffuse radio emission, which
is not identified with any individual galaxy and 
makes the Coma cluster exceptional in the radio domain. 
Cluster-wide diffuse sources 
witness the existence of relativistic electrons and large
scale magnetic field, associated with the intergalactic medium.
They are known to be present so far only in
a few clusters, despite many efforts to detect them. 
%They are
%classified in the literature in  two classes: radio halos, present at the 
%cluster center, and relics, located at the cluster periphery.
The diffuse radio emission in Coma includes 
%both classes of sources: 
the central radio source Coma C, the peripheral
diffuse source 1253+275, and the bridge of very low brightness between 
them. A radio image of the Coma cluster is presented in Fig. 1.
A Hubble constant H$_0$ = 50 km s$^{-1}$ Mpc$^{-1}$ is used throughout.
\par\noindent
\underline {The radio galaxies}:
There are 29 cluster galaxies~\cite{ve} showing radio emission down to a 
power level of 
 $\sim$10$^{21}$ W Hz$^{-1}$: 
10 are Ellipticals, 3 S0 + E/S0, and 16 Spirals + Irregulars.
Among the extended radio galaxies identified with ellipticals, the dominant 
radio structure is the ``tailed'' structure, typically found in clusters,
and determined by the drag exerted by the intracluster gas  on the 
radio emitting plasma ejected from moving galaxies.
In particular, the giant central galaxy NGC~4874 is 
known to be associated with a small ($\sim$15 kpc)
tailed radio source~\cite{fg}, which indicates that the
galaxy is not stationary at the cluster center. Also the giant
peripheral galaxy NGC~4839 shows a tailed radio morphology~\cite{ve}, 
with the low brightness tails oriented away from the cluster center.
A prominent tailed radio source, extended about 300 kpc,
 is  associated with NGC~4869~\cite{fd}, located at $\sim$4$^{\prime}$ from 
NGC~4874. Finally, the other tailed
radio galaxies in Coma are NGC~4849~\cite{ve}, and NGC~4789~\cite{vg}, which
lie in the cluster periphery, at $\sim$1.5$^{\circ}$ from
the center.
\par\noindent
\underline{The radio halo Coma C}:
Coma C is the prototype and best studied example of cluster radio 
halos~\cite{gf}.
It is at the cluster center, shows a rather regular shape, a size 
of $\sim$ 1.1 Mpc, a low surface brightness ($\sim$.1 mJy at 1.4 GHz),
and steep radio spectrum ($\alpha$=1.34). No polarized flux is
detected down to a level of $\sim$10\% at 1.4 GHz.
The halo Coma C is the only radio halo, for which a high resolution
map of the spectral index
has been obtained so far. The spectral index 
distribution shows a central plateau 
with $\alpha \sim$0.8, and an outer region with steeper spectrum, up
to $\alpha$=1.8. This behaviour provides evidence that the source of energy
is more efficient at the cluster/halo center, in a region approximately 
coincident with the optical core radius.
The  radiative lifetime of the relativistic electrons  
estimated from the spectrum is $\sim$10$^8$yr.
Recently, Deiss et al.~\cite{de}  obtained a map at 1.4 GHz of the halo, after
subtraction of all discrete sources. They pointed out the close similarity
between the X-ray and radio map: similar extension both 
in the E-W direction and
toward  the NGC~4839 group. This similarity indicates a close link between
the physical conditions of the radio source and those of the thermal component.
\par\noindent
\underline{The relic 1253+275}:
The peripheral  diffuse source 1253+275~\cite{gs}, is at  $\sim$2.7 Mpc
from the cluster
center, in the direction of the cluster A1367.
It is classified as a relic
source, as it has been suggested to be the remnant of the radioemission of a
currently inactive radio galaxy,  but no compelling
evidence  of this has been found so far. Its largest size,
brightness and spectrum are
similar to those of Coma C. Unlike Coma C it shows an elongated shape, 
and is polarized at the 30\% level at 1.4 GHz. 
The high polarization degree in this source is naturally explained if
a tangled magnetic field is associated with the cluster intergalactic medium.
In this case, a larger number of magnetic field cells along the line of
sight is present at the cluster center compared to the outer regions.
\par\noindent
\underline{The bridge}:
A bridge of radio emission~\cite{kk} is present in the region connecting
Coma C to 1253+275. The surface brightness of this diffuse
emission is very low, and is only enhanced at low frequency and low
resolution. From the  recent map presented 
by Deiss et al.~\cite{de}, it
is questionable if the bridge is a feature connecting Coma C to 1253+275,
or an asymmetric extension of the central halo Coma C.

\section{The magnetic field in Coma}

The value of the equipartition magnetic field in the diffuse radio sources
of Coma is $\sim$0.4 $\mu$G. A much larger value is obtained with an 
indipendent argument, by
studying the polarization properties  of the radio galaxy NGC~4869~\cite{fa},
embedded within the intergalactic medium 
in the central cluster region.
The fluctuations in the rotation measure detected in this source
imply that the magnetic field is tangled on scales
of about 1.5 kpc,  and has a strength of  $\sim$6 $\mu$G.
%This result is consistent with that obtained from the statistical
%analysis of Kim et al. (1991),
%if the different value of the magnetic field tangling scale is taken into
%account. 
Such a large magnetic field has important implications on the energetics of 
the radio halo Coma C, and implies the existence of amplification mechanisms.

\section{Origin of the diffuse emission}

The relativistic electrons radiating in the diffuse radio emission 
through  the synchrotron process, 
can travel at most a distance of $\sim$100 kpc over their
radiative  lifetime ($\simeq$ 
10$^8$ yr), with a diffusion velocity  of $\sim$1000 km/s
(the typical sound speed).
 This implies that ``in situ'' reacceleration of the relativistic
electrons is needed to mantain the radio emission 
on a scale of more than 1 Mpc.
It has been suggested that a recent merger between subgroups could provide the
energy necessary to maintain the radio halos, i.e. 
 reaccelerate the radiating electrons and  amplify the
magnetic field (see e.g. the review by Feretti \& Giovannini~\cite{bo}).  
B\"ohringer et al.~\cite{ha} estimated that typically a
power of the order of  3 $\times$ 10$^{45}$ erg s$^{-1}$ is 
available from a merger. This
is enough to power the halo with reasonably low efficiency.

According to this scenario, the merger in the center of Coma, between
the groups of NGC~4874 and NGC~4889~\cite{co} can be responsible for 
the energy supply to Coma C, while
the merger between the NGC~4839 group and the main cluster~\cite{br}, could
play a role in the energy supply to the bridge. 
The possibility that the group of NGC~4839 has 
already passed through the cluster core $\sim$2 Gyrs ago~\cite{bu}
does not seem  to be relevant for the formation of the radio halo Coma C,
since it is difficult to reconcile the timescale of the merger event
to the electron lifetime and to the spectral behaviour of Coma C.
We do not see any connection between the peripheral relic source 
1253+275 and a cluster merger. This could favour the possibility suggested
in the literature~\cite{gs} that this is the remnant of a radio galaxy.

The connection between a merger process and the presence of a diffuse 
halo does not explain the rarity of radio halos, as mergers
are relatively common in clusters of galaxies. According to 
Giovannini et al.~\cite{gf},
the rarity of radio halos could be due to
the difficulty in obtaining the relativistic electrons radiating
within the radio halo. These authors suggested that tailed 
radio galaxies orbiting at the cluster center may be responsible 
for the deposit of relativstic electrons  and estimated that 
the number of particles radiating in Coma C can be supplied
by NGC~4869 in about 4 orbits.
Alternatively, Deiss et al.~\cite{de} suggested that the 
efficiency of the stochastic reacceleration mechanism operating in the 
intracluster medium can explain the 
close connection between radio emission and X-ray gas structure,
and the rarity of radio halos. 
They argue that the reacceleration by turbulent gas motion originated 
by galaxy motions may depend  on some details of the cluster structure, so that
only in a few clusters the electrons gain enough energy to produce an
observable halo. This process seems
to be sufficiently strong to account for the emission in Coma C and to
sustain the cluster-wide distribution of relativistic particles.

Therefore, the complex diffuse radio emission in Coma 
is not yet fully understood, and in particular the possibility that 
the merger process is not the only ingredient in its formation should be
considered.

\section {Conclusions}

The Coma cluster is permeated by diffuse radio emission, which is a rare
phenomenon in clusters of galaxies. 
It seems that merger processes can supply energy for the maintenance
of the central radio halo, and the bridge, while the origin of the
peripheral source 1253+275 is still puzzling.
Models of dynamical evolution of the Coma cluster must account for the
presence of the diffuse radio emission.

%\section*{Acknowledgments}

\begin{figure}[t]
\includegraphics{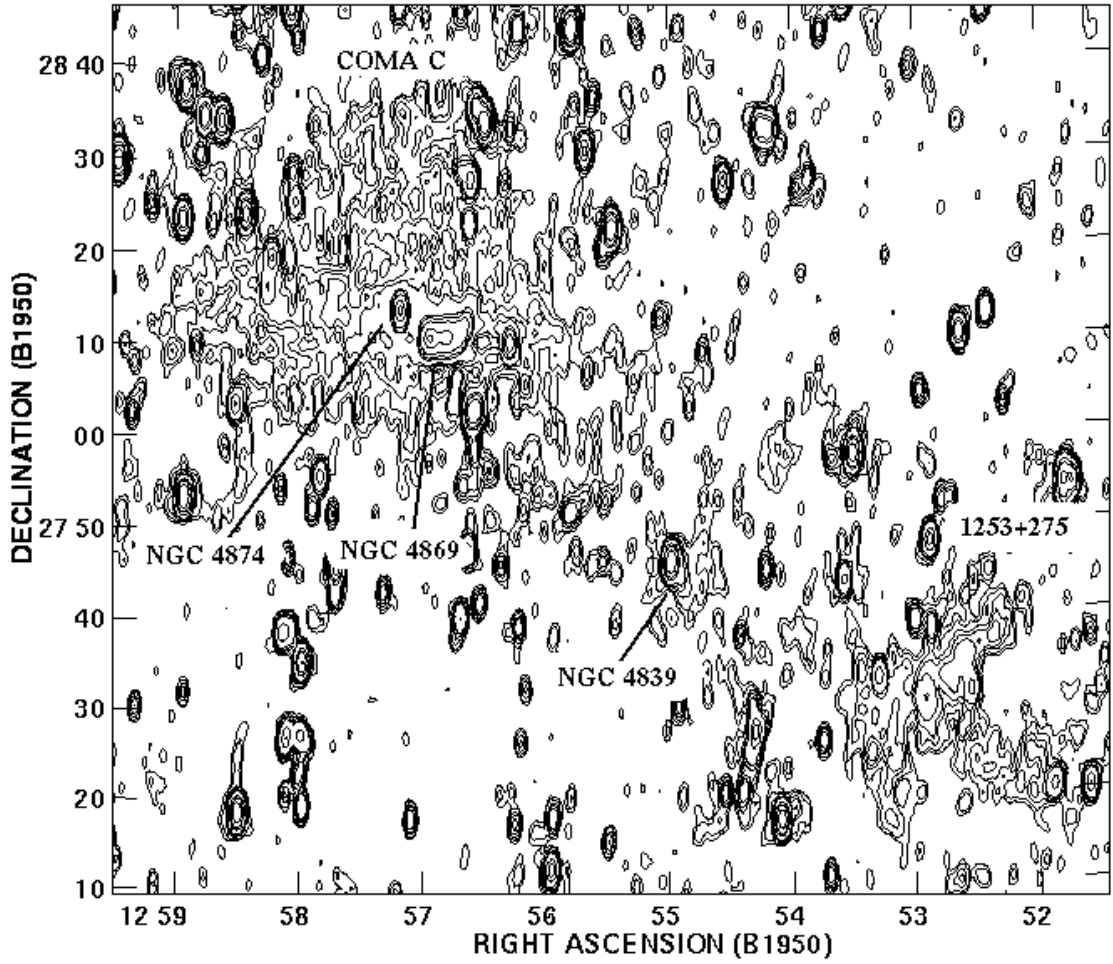}
\vspace{13 cm}
\caption{Radio map of Coma, obtained with the Westerbork Synthesis Radio 
Telescope at 90 cm. The angular resolution is 51$^{\prime\prime}$ $\times$ 
122$^{\prime\prime}$ (HPBW, RA $\times$ DEC). Contour levels 
are 2, 3, 5, 7, 10, 30, 50, 100, 300, 500 mJy/beam. The central halo Coma C
is clearly visible, as well as the peripheral diffuse source 1253+275. The
bridge between them is barely visible in this image, because of its very low
brightness. }
\end{figure}

\end{document}